\documentclass[a4paper]{jpconf}
\usepackage{graphicx}
\begin{document}
\title{Hybrid method for identifying mass groups of primary cosmic rays in the joint operation of IACTs and wide angle Cherenkov timing arrays}

\author{E B Postnikov$^{1,*}$, A A Grinyuk$^2$,
	L A Kuzmichev$^1$ and L G Sveshnikova$^1$}
\address{$^1$ Lomonosov Moscow State University Skobeltsyn Institute of Nuclear Physics (MSU SINP), Leninskie gory 1(2), GSP-1, Moscow, 119991, Russia}
\address{$^2$ Joint Institute for Nuclear Research (JINR), Joliot-Curie 6, Dubna, Moscow region, 141980, Russia}

\ead{$^*$evgeny.post@gmail.com\\
}

\begin{abstract}
This work is a methodical study of another option of the hybrid method originally aimed at gamma/hadron separation in the TAIGA experiment. In the present paper this technique was performed to distinguish between different mass groups of cosmic rays in the energy range 200 TeV -- 500 TeV. The study was based on simulation data of TAIGA prototype and included analysis of geometrical form of images produced by different nuclei in the IACT simulation as well as shower core parameters reconstructed using timing array simulation. We show that the hybrid method can be sufficiently effective to precisely distinguish between mass groups of cosmic rays.
\end{abstract}

\section{Introduction}
The measurement of the mass composition of cosmic rays could be the key to understanding their origin, because the change of dominant sources (mostly supernova remnants of different types) in the given energy range should lead to the corresponding change of mass composition \cite{1, 2}. The energy range of hundreds of TeV is of special interest, because the most frequent in our Galaxy supernova remnants of type II can probably accelerate cosmic rays up to about 100 TeV \cite{1}.
 
However, the mass composition of cosmic rays in this energy range is poorly measured and remains uncertain because neither type of experiment can achieve it: satellite-borne cosmic ray detectors applicable for this range have a very limited geometric factor, whereas standard methods of extensive air shower measurement are not suited for so low energy region. In this paper we consider a possibility to distinguish between different mass groups of cosmic rays in the energy range 200--500 TeV using the hybrid method of Cherenkov light registration originally aimed at gamma/hadron separation in the TAIGA experiment. The observatory TAIGA (Tunka Advanced Instrument for cosmic ray physics and Gamma-ray Astronomy \cite{3}) is designed for high energy gamma ray ($>$30 TeV) and cosmic ray ($>$100 TeV) measurements. It combines the cost-effective wide angle timing array with a set of IACTs (Imaging Atmospheric Cherenkov Telescopes) to allow reaching a total area up to a few square kilometers and strong suppressing hadron background \cite{3}.

To the best of our knowledge, the mass group identification with IACT has never been realized yet, even though IACT images are sensitive to the depth of shower maximum and therefore also to the mass of primary particles. The only exception we know is a spectrum of iron nuclei in the interval 13--200 TeV derived by HESS \cite{4}. However, their method relies on the ground based detection of Cherenkov light emitted by the primary particle prior to its first interaction in the atmosphere. The other example can be referred to \cite{5}, where a hybrid analysis was carried out by combining the data coming from ARGO-YBJ and a wide field of view Cherenkov telescope WFCT. The latter wasn't a standard IACT because it didn't have mirror, but only a set of $16\times16$ photomultipliers (PMTs) with a large field of view $16^{\circ}\times16^\circ$ and a pixel size of approximately $1^\circ$. The light component spectrum (p+He) in the 100 TeV -- 3 PeV energy region was measured after the hybrid selection technique \cite{5, 6}. Nevertheless, the application of the real contemporary IACT technique to solving this task is yet to come. In this paper we discuss the ability to distinguish between different mass groups of cosmic rays in the TAIGA hybrid experiment.

\section{Monte Carlo simulations}
\label{sec-2}
Simulation was performed for primary protons and nuclei with the fixed energy 200 and 500 TeV incident within $\pm5^\circ$ around the fixed pointing direction of the IACT (zenith angle 30$^\circ$). This configuration is close to the TAIGA prototype design. In the present work we simulated in details only the images produced by showers in a camera of an IACT. The arrival direction, energy and core position of every shower were assumed to be known with the accuracy $\pm0.1^\circ$ and $\pm$10~m correspondingly. Such resolutions are typical for the HISCORE timing array \cite{3}.

At the first step, the shower development in the atmosphere was simulated with the CORSIKA package \cite{7}. The response of the IACT system was simulated at the second step using our software developed for this task. The segmented mirror of an IACT has an area of about 10 m$^2$ and a focal length of 4.75 m. The camera located at the focus consists of 560 photomultipliers with the total field of view $9.72^\circ$ (FOV), and the single pixel FOV is $0.36^\circ$. Cherenkov photons of the shower were traced through the IACT optical system and the number of corresponding photoelectrons in each pixel of the camera was counted. 

\section{Methods}
\label{sec-3}
\subsection{Quality criterion}
As in the case of gamma ray selection \cite{8}, for each event from data bank various features of the image were calculated. The purpose of the study was to determine the most distinctive feature of nuclei and proton/helium induced images. As a quality criterion of particle separation the selection quality factor $Q$ was estimated. This factor indicates an improvement of a significance $S_0$ of the statistical hypothesis that the events do not belong to the background. For Poisson distribution (that is for a large number of events), the selection quality factor is:
\begin{equation}
Q=\epsilon_{nuclei}/\sqrt{\epsilon_{bckgr}},
\end{equation}
where $\epsilon_{nuclei}$ and $\epsilon_{bckgr}$ are relative numbers of selected events and background events after selection. For our task we consider p+He as a background and try to select oxygen and iron nuclei above this background. The optimal image features for the selection were found in a process of maximizing $Q$ value under the condition $\epsilon_{nuclei}\geq0.5$. 
\subsection{Night sky background reduction}
In experiment an image is distorted by the night sky background (NSB) following Poisson distribution. To obtain a more reliable estimation of $Q$ in simulation, the NSB was randomly replicated for every image and $Q$ was averaged over all replications. The procedure of image reconstruction above the NSB is called image cleaning \cite{9}. Our cleaning procedure was developed as to maximize quality of subsequent selection nuclei/background. All the results below were obtained after applying the optimal procedure of image cleaning. 

\subsection{Specific features of the problem}
The IACT procedure of nuclei discrimination against p/He may have specific features. 

Unlike gamma ray showers, air showers induced by primary nuclei do not arrive from the point source, but their directions are isotropic. For that reason an event preselection inside the narrow cone around the true source direction is not allowable. This preselection was the additional advantage of the hybrid technique and allowed us to sufficiently reduce background and calculate the selection quality only for the narrow space cone $0.1^\circ\times0.1^\circ$ \cite{8}. Furthermore, the fact that the telescope is not pointed at the origin of the events to be selected makes also useless very effective selection algorithms. For example, the Hillas parameter 'azimuthal width', which proved the best one for gamma ray selection in TAIGA simulation \cite{8}, is worthless for nuclei selection. This parameter is a width of an image ellipse relative to an axis between the image center and the center of the camera \cite{10}. Gamma ray induced images are pointed at the camera center, which is the center of the FOV, because of the telescope orientation at their source. Unfortunately, nuclei have the same isotropic distribution as proton/helium background.

However, the advantage of the selection of nuclei is that we do not have to reach as great rejection power as for gamma rays because the latter are much rarer than the former. Therefore, the rejection of one order of magnitude ($Q\sim2$) would be sufficient for scientific purposes. 

The second positive point of nuclei selection by a hybrid technique is the use of another type of 'image projection' parameter, which was introduced by authors of the present paper in \cite{8}. This group of parameters is based on the shower core position and is calculated in a way suggested by Hillas for azimuthal width but implemented not for the axis 'image center - center of the camera' but for the 'image center - core position in a camera plane'. Hybrid imaging/timing experiment setup and corresponding imaging/timing technique of data analysis assume that the core position is to be determined with a great accuracy (in TAIGA $\sim$5--6 m \cite{3}) and therefore the data for this new parameter calculation will be provided.

\subsection{Image features calculation}
\label{sec-3.3}
We calculated various image characteristics potentially useful for discrimination between nuclei and protons/helium. They could be sorted out into 2 groups:
\begin{itemize}
\item Hillas parameters \cite{10}, unrelated to data projection on some definite direction;
\item `core-azimuthal' parameters, introduced in the present paper and in \cite{8}.
\end{itemize}
The 3rd group of parameters, `azimuthal' width and length \cite{10}, can not be used for nuclei/background discrimination because of absence of a certain arrival direction of primary nuclei.

\begin{figure}[h]
\begin{minipage}{12pc}
\includegraphics[width=12pc]{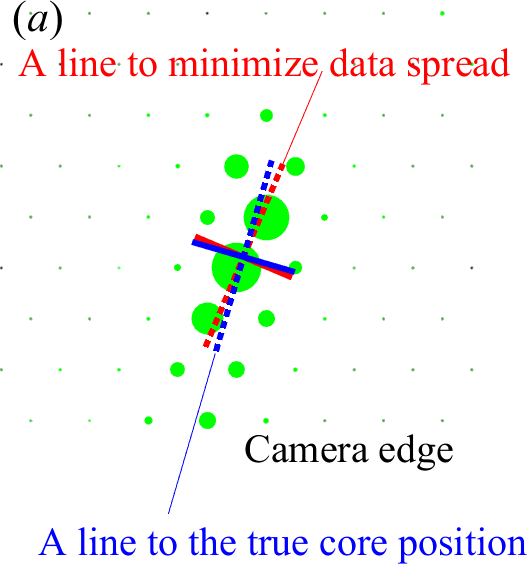}
\end{minipage}
\begin{minipage}{12pc}
\includegraphics[width=12pc]{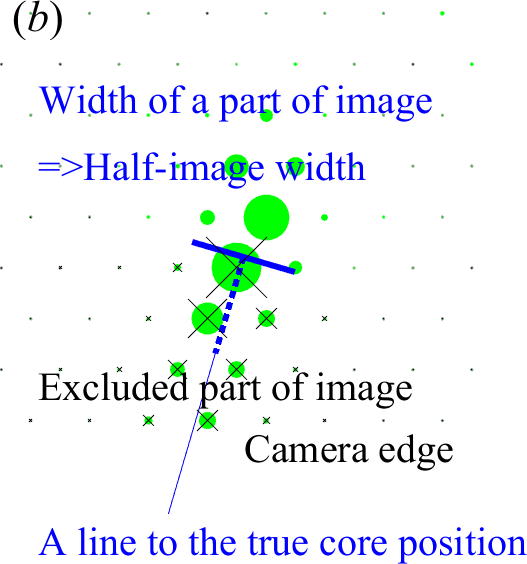}
\end{minipage} 
\begin{minipage}{13pc}
\caption{\label{figure1}Image features computation: (\textit{a}) length and width (red \dotted and thick \full), core-azimuthal length and width (same blue lines); (\textit{b}) core-azimuthal half-image width.}
\end{minipage} 
\end{figure}

The principle of calculation of parameters from both groups is illustrated in figure \ref{figure1}. For the 1st group, the image axis is determined as a line minimizing the weighted sum of squares of a distance to the pixels. Then the width and length can be calculated as 2nd central momenta of image intensity distribution with respect to this axis. For the 2nd group, the image axis is to be drawn through the center of gravity of the image to connect it with the known shower core position. The core position was randomized with $\sim$10 m accuracy (section \ref{sec-2}). 

For both groups we determined also kurtosis (the 4th momentum divided by squared variance) and half-image width parameter designed to account for image truncation in case of distant showers. The half-image width is introduced in the present paper especially for the hybrid technique conditions, because no image truncation is allowable in standard IACT analysis when the distance from the shower core to the IACT is much less than in the TAIGA experiment \cite{11, 3}.

\section{Results}
\label{sec-4}
\begin{figure}[h]
\begin{minipage}{24pc}
\includegraphics[width=24pc]{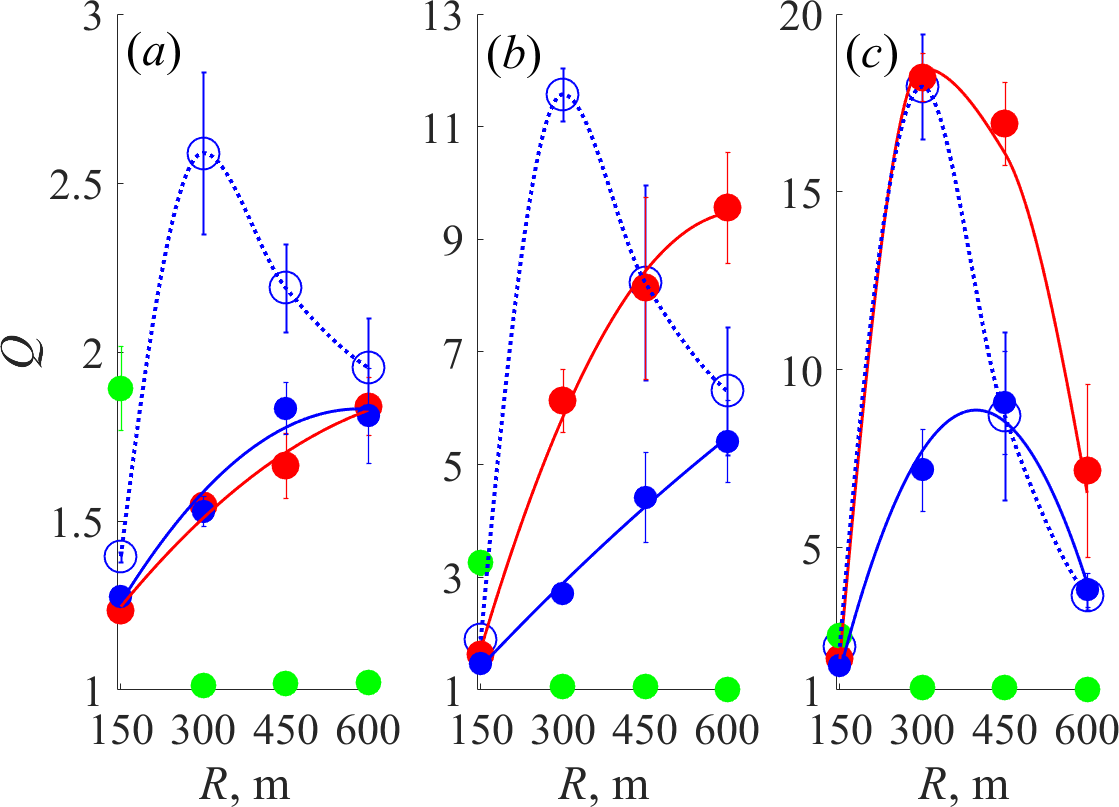}
\end{minipage}
\begin{minipage}{14pc}
\caption{\label{figure2}Selection quality factor vs shower core distance. (\textit{a})~O~500~TeV, (\textit{b})~Fe~500~TeV, (\textit{c})~Fe~200~TeV. Green \fullcircle -- image kurtosis, red \fullcircle -- image width, blue \fullcircle -- half-image width, blue \opencircle -- core-azimuthal half-image width. Solid and dotted lines are guide to the eyes.}
\end{minipage}
\end{figure}

\begin{figure}[h]
\begin{minipage}{15pc}
\includegraphics[width=15pc]{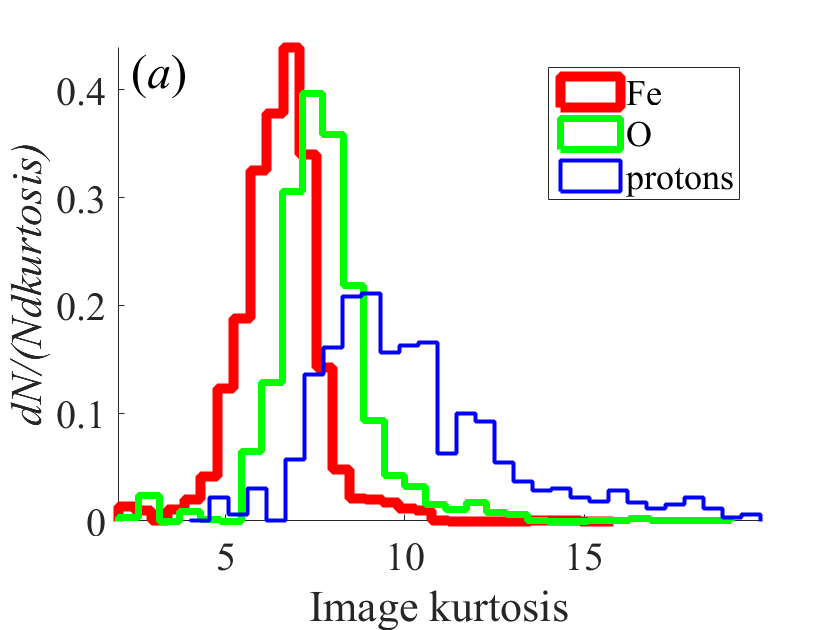}
\end{minipage}
\begin{minipage}{15pc}
\includegraphics[width=15pc]{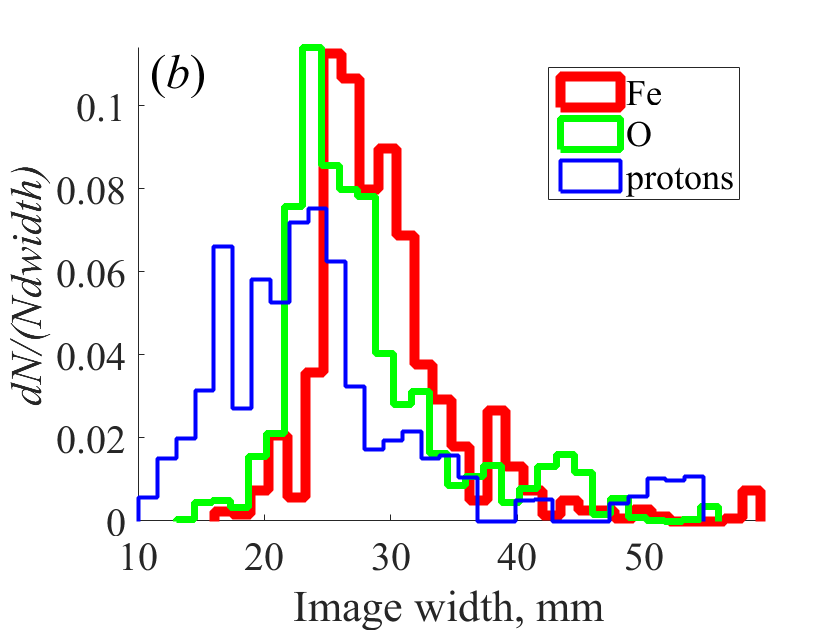}
\end{minipage} 
\begin{minipage}{7pc}
\caption{\label{figure3}Discriminating parameter histogram: (\textit{a}) kurtosis and (\textit{b}) width for 0--150~m, (\textit{c}) core-azimuthal half-image width and (\textit{d}) width for 150--300~m. $E$=500~TeV.}
\end{minipage} 
\begin{minipage}{15pc}
\includegraphics[width=15pc]{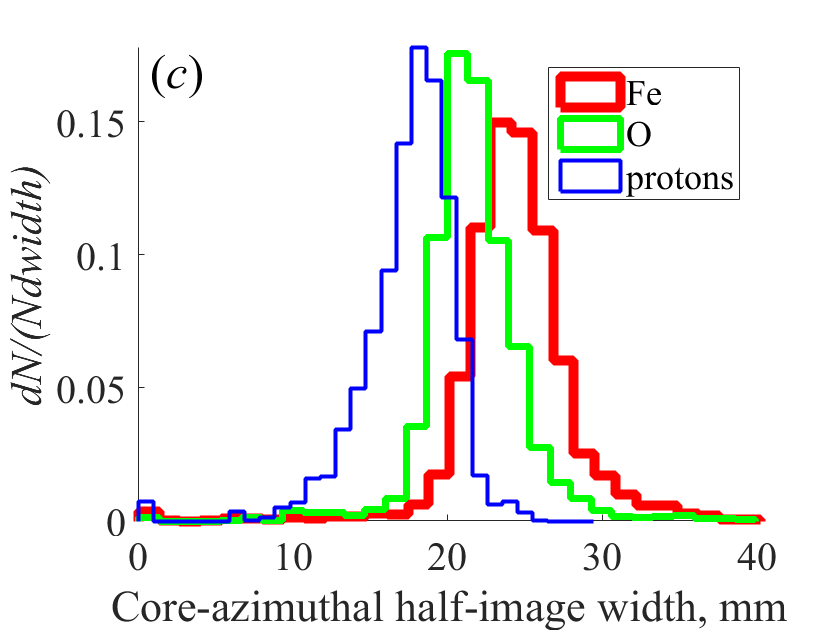}
\end{minipage}
\begin{minipage}{15pc}
\includegraphics[width=15pc]{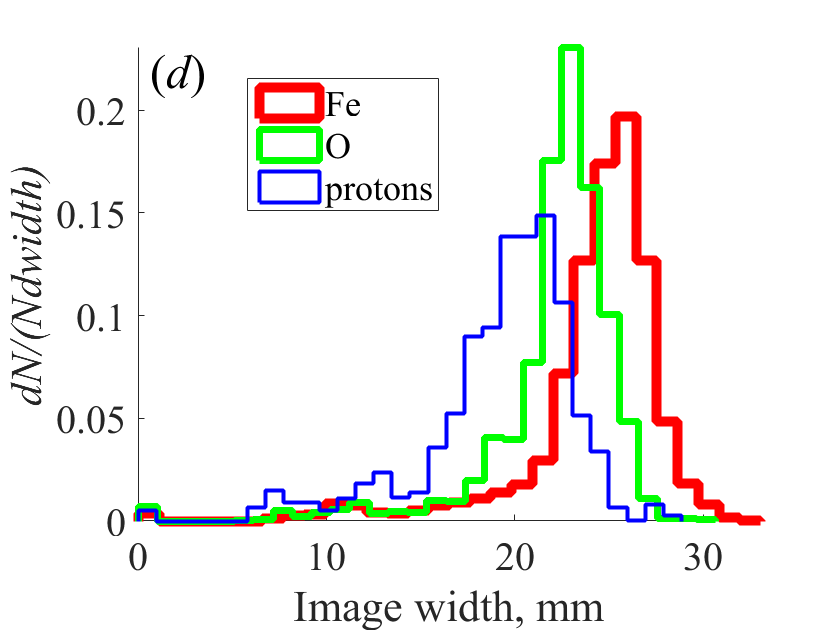}
\end{minipage} 
\end{figure}

Values of selection quality $Q$ for the three best parameters are plotted in figure \ref{figure2} for Fe and O nuclei, 200--500 TeV. For small distance $R$ from the shower to the telescope the kurtosis proved the best, whereas for greater distance the best was the parameter from the 2nd group (section \ref{sec-3.3}), the core-azimuthal half-image width (except 200 TeV, see the figure). Comparison of these parameters distribution with the standard width of an image is presented in figure \ref{figure3}.
\section{Discussion}
\label{sec-5}
As for gamma rays \cite{8}, the best discrimination was achieved for $R\sim$200--300 m, however, the best parameter for this region is the half-image width (from the group of core-azimuthal parameters). This can be explained by image truncation starting with this distance. For small distance the best is a kurtosis describing an image intensity distribution form rather than its dimensions.

\section{Conclusions}
\label{sec-6}
Our methodical study revealed sufficient quality of discrimination of air showers induced by primary nuclei against background (primary proton/helium showers). This kind of discrimination can be obtained in a hybrid timing/imaging array using combination of an imaging Cherenkov air telescope and timing detectors. Such hybrid timing/imaging array is to be realized in the Tunka valley in 2017 as a gamma ray observatory, however, the results of the present study give it an option of successful work with nuclei. All necessary modifications of the selection technique are described. The proton/helium background suppression about 100 times can be achieved for Fe nuclei and about 10 times for oxygen at a distance up to 600 m and energy 200--500 TeV. The influence of combining both types of data, imaging and timing, on discrimination quality is stronger than for gamma ray selection and therefore this non-standard use of imaging telescope can also be realized in the framework of a hybrid technique.
\ack
The study was supported by the Russian Foundation for Basic Research, project 16-29-13035.

\end{document}